\global\def\draftcontrol{0}

   \def\versionno{ bulk oz et.al }

\catcode`\@=11

\expandafter\ifx\csname draftcontrol\endcsname\relax\global\def\draftcontrol{0}
\fi

{\count255=\time\divide\count255 by 60
\xdef\hourmin{\number\count255}
\multiply\count255 by-60\advance\count255 by\time
\xdef\hourmin{\hourmin:\ifnum\count255<10 0\fi\the\count255}}
\def\draftdate{\number\month/\number\day/\number\year\ \ \ \hourmin }

\newcommand\makepapertitle{\par
  \begingroup
    \renewcommand\thefootnote{\@fnsymbol\c@footnote}%
    \def\@makefnmark{\rlap{\@textsuperscript{\normalfont\@thefnmark}}}%
    \long\def\@makefntext##1{\parindent 1em\noindent
            \hb@xt@1.8em{%
                \hss\@textsuperscript{\normalfont\@thefnmark}}##1}%
     \newpage
     \global\@topnum\z@   
     \@makepapertitle
     \thispagestyle{empty}\@thanks
  \endgroup
  \setcounter{footnote}{0}%
  \global\let\thanks\relax
  \global\let\makepapertitle\relax
  \global\let\@makepapertitle\relax
  \global\let\@thanks\@empty
  \global\let\@author\@empty
  \global\let\@date\@empty
  \global\let\@title\@empty
  \global\let\title\relax
  \global\let\author\relax
  \global\let\date\relax
  \global\let\and\relax
  \def\version{\let\version\@version\@gobble}
}
\def\@makepapertitle{%
  \newpage
   \ifnum\draftcontrol=1 {}
   \version\versionno
   \vskip 3em%
   \else
   \hfill\hbox to 3cm {\parbox{4cm}{\@pubnum}\hss}%
   \vskip 3em%
   \fi
   \begin{center}%
   \let \footnote \thanks
     {\LARGE {\@title}}%
     \vskip 1.5em%
     {\normalsize
       \lineskip .5em%
       \begin{tabular}[t]{c}%
         \@author
       \end{tabular}\par}%
     \vskip 1.5em%
     {\@bstract}%
     \end{center}%
     \vskip 1.5em
     \@date%
   \par
}

\gdef\@pubnum{}
\def\pubnum#1{%
  \gdef\@pubnum{#1}}

\gdef\@bstract{}
\def\Abstract#1{%
  \gdef\@bstract{%
   \parbox{\textwidth-0pc}{%
   \centerline{\bf Abstract}\penalty1000%
\kern.2cm%
\noindent
\renewcommand\baselinestretch{1.0}%
{#1}}}
}

\def\ps@paper{\let\@mkboth\@gobbletwo%
     \ifnum\draftcontrol=1
    \def\@oddfoot{\hbox to \textwidth{\tiny \versionno \hfil\tiny\draftdate}%
    \hskip -\textwidth \hbox to \textwidth{\hfil\rm\thepage\hfil}}%
     \else\def\@oddfoot{\hbox to \textwidth{\hfil\rm\thepage\hfil}}
     \fi
     \let\@evenfoot\@oddfoot
}

\def\body{\clearpage
          \pagestyle{paper}
    }

\def\@version#1{\ifnum\draftcontrol=1
\typeout{}\typeout{#1}\typeout{}
\vskip3mm\centerline{\hbox{\fbox{\normalsize{\tt DRAFT -- #1 -- }
                   {\draftdate}}}}\vskip3mm
\fi}
\let\version\@version
\long\def\eqlabel#1{\ifnum\draftcontrol=1
                    \tag@false  
                    \tag*{(\theequation) \hbox to -0.2cm{\hspace{0cm}\small{#1}\hss}}
                    \refstepcounter{equation}
                    \edef\@currentlabel{\theequation}
                    \ltx@label{#1}          
                    \else
                    \label{#1}
                    \fi
                    }
\let\st@bibitem\@bibitem
\let\st@lbibitem\@lbibitem
\ifnum\draftcontrol=1
  \def\@bibitem#1{%
    \st@bibitem{#1}\a@@label{#1}\ignorespaces}
  \def\@lbibitem[#1]#2{%
    \st@lbibitem[#1]{#2}\a@@label{#2}\ignorespaces}
  \def\a@@label#1{%
    \gdef\a@lab{\smash{\normalfont\small#1}}
    \ifvmode
      \if@inlabel
        \global\setbox\@labels\hbox{%
          \llap{\a@lab\let\a@lab\relax
                \kern\@totalleftmargin\kern\marginparsep}%
          \box\@labels}%
      \fi
    \fi}
\fi

\documentclass[12pt,letterpaper]{article}

\usepackage{amsmath,amssymb,array,calc,epsfig,rotating,bm}
\usepackage[sort]{cite}
\usepackage{graphicx}
\usepackage{psfrag,verbatim}

\ifnum\draftcontrol=1
\fi

\renewcommand\baselinestretch{1.25}
\renewcommand\section{\@startsection {section}{1}{\z@}%
                                   {-3.5ex \@plus -1ex \@minus -.2ex}%
                                   {2.3ex \@plus.2ex}%
                                   {\normalfont\large\bfseries}}
\renewcommand\subsection{\@startsection{subsection}{2}{\z@}%
                                   {-3.25ex\@plus -1ex \@minus -.2ex}%
                                   {1.5ex \@plus .2ex}%
                                   {\normalfont\normalsize\bfseries}}
\renewcommand\subsubsection{\@startsection{subsubsection}{3}{\z@}%
                                   {-3.25ex\@plus -1ex \@minus -.2ex}%
                                   {1.5ex \@plus .2ex}%
                                   {\normalfont\normalsize\it}}
\renewcommand\paragraph{\@startsection{paragraph}{4}{\z@}%
                                   {-3.25ex\@plus -1ex \@minus -.2ex}%
                                   {1.5ex \@plus .2ex}%
                                   {\normalfont\normalsize\bf}}

\numberwithin{equation}{section}

\def\revise#1       {\raisebox{-0em}{\rule{3pt}{1em}}%
                     \marginpar{\raisebox{.5em}{\vrule width3pt\
                     \vrule width0pt height 0pt depth0.5em
                     \hbox to 0cm{\hspace{0cm}{%
                     \parbox[t]{4em}{\raggedright\footnotesize{#1}}}\hss}}}}

\def\cale         {{\cal E}}

\def\calm         {{\cal M}}
\def\caln         {{\cal N}}
\def\calo         {{\cal O}}
\def\calp         {{\cal P}}

\def\calr         {{\cal R}}

\def\calv         {{\cal V}}

\def\del          {\partial}

\def\sqr#1#2{{\vcenter{\vbox{\hrule height.#2pt
 \hbox{\vrule width.#2pt height#1pt \kern#1pt
 \vrule width.#2pt}\hrule height.#2pt}}}}

\newcommand{\ft}[2]{{\textstyle{\frac{#1}{#2}}}}

\def\a{\alpha}
\def\b{\beta}

\newcommand{\beq}{\begin{equation}}
\newcommand{\eeq}{\end{equation}}
\newcommand{\beqa}{\begin{eqnarray}}
\newcommand{\eeqa}{\end{eqnarray}}
\newcommand{\beqar}{\begin{eqnarray*}}
\newcommand{\eeqar}{\end{eqnarray*}}

\renewcommand{\eqref}[1]{(\ref{#1})}

\newcommand{\ie}{{\it i.e.,}\ }

\def\a{\alpha}
\def\w{\omega}

\def\r{\rho}

\def\dd{{\delta}}
\def\l{\lambda}

\def\k{\kappa}

\def\ta{\tilde{a}}
\def\ha{\hat{a}}
\def\z{\zeta}
\catcode`\@=12

\begin{document}


\title{\bf On Eling-Oz formula for the holographic bulk viscosity}
\pubnum
{UWO-TH-11/4
}

\date{March 2011}

\author{
Alex Buchel\\[0.4cm]
\it Department of Applied Mathematics\\
\it University of Western Ontario\\
\it London, Ontario N6A 5B7, Canada\\
\it Perimeter Institute for Theoretical Physics\\
\it Waterloo, Ontario N2J 2W9, Canada\\
}

\Abstract{
Recently Eling and Oz \cite{eo} proposed a simple formula for the bulk
viscosity of holographic plasma.  They argued that the formula is
valid in the high temperature (near-conformal) regime, but is expected
to break down at low temperatures. We point out that the formula is in
perfect agreement with the previous computations of the bulk viscosity
of the cascading plasma
\cite{c1,c2}, as well as with the previous computations of the bulk viscosity of  
$\caln=2^*$ plasma \cite{n21,n22}.
In the latter case it correctly reproduces the critical behaviour of
the bulk viscosity in the vicinity of the critical point with the
vanishing speed of sound.
}

\makepapertitle

\body

\version\versionno
\tableofcontents

\section{Introduction and Summary}
In \cite{eo} Eling and Oz (EO) considered\footnote{See \cite{fujita} for 
related earlier work.} 
effective five-dimensional gravitational description of 
strongly coupled gauge theory plasma
frequently arising both in phenomenological 
and the formal examples of the holographic gauge theory/string theory correspondence \cite{juan,adscft}
\begin{equation}
S_5=\frac{1}{16 G_5\pi} \int d^5 x\ \sqrt{-g}\left(R_5-\frac12\sum_i\left(\del\phi_i\right)^2-V(\phi_i)\right)
+S_{gauge}\,.
\eqlabel{5daction}
\end{equation}
In the absence of chemical potentials for the conserved $U(1)$ charges they proposed a simple 
formula for the ratio of bulk-to-shear viscosities in holographic plasma dual to \eqref{5daction}:
\begin{equation}
\frac{\zeta}{\eta}=\sum_i\ c_s^4 T^2 \left(\frac{d\phi_i^H}{dT}\right)^2\,,
\eqlabel{zeta}
\end{equation}
where $T$ is the plasma temperature (equivalently the Hawking temperature of the black brane 
dual to the equilibrium thermal state of the plasma), 
\begin{equation}
c_s^2=\frac{d(\ln T)}{d(\ln s)}\,,
\end{equation}
is the speed of sound waves in plasma, and $\phi_i^H$ are the values of the gravitational  scalar fields 
$\phi_i$  evaluated at the black brane horizon. The remarkable feature of the expression 
\eqref{zeta} is in the fact that it can be evaluated entirely from the 'horizon data'. 
The reason why this is unexpected is that the bulk viscosity is sensitive to the microscopic 
scales in the theory, which in holographic correspondence  are encoded in non-normalizable components of the 
appropriate scalar fields near the boundary\footnote{This should be contrasted with the 
familiar universality of the shear viscosity to the entropy density ratio \cite{u1,u2,u3,u4} which 
can be evaluated in the membrane paradigm framework \cite{m1}.}.  
In fact, above was precisely the reason the authors of \cite{eo} suggested that the 
formula \eqref{zeta} validity should be restricted to the high temperature case --- in this 
case the scalar fields are expected not to flow a lot from the boundary to the horizon, and thus 
their horizon values should remain sensitive to the microscopic parameters of the boundary 
gauge theory.  EO formula \eqref{zeta} was successfully verified \cite{eo} for 
a wide set of gauge theory/string theory examples, where bulk viscosity was evaluated 
from the sound waves attenuation coefficient \cite{v1,v2,v3,v4}. Furthermore, it was found \cite{eo} that 
while \eqref{zeta} is valid in some phenomenological models of gauge/gravity correspondence \cite{vp1,vp2,f2}, 
it is violated in some other models  \cite{f1,f3}. 

In this paper we demonstrate that, at least in the context of formal examples of holographic 
correspondence, the range of validity of \eqref{zeta} is much wider. In particular, 
in section 2 we verify \eqref{zeta} for the cascading gauge theory plasma to 
order $\calo\left(\ln^{-4}\frac T\Lambda\right)$ in the high-temperature $T\gg \Lambda$ 
expansion\footnote{Original computations of the bulk viscosity in this theory were done in \cite{c1,c2}.}.
Results are presented in section 2.1.
In section 3 we verify \eqref{zeta} for $\caln=2^*$ gauge theory plasma for all 
temperatures\footnote{Original computations of the bulk viscosity in this theory were done in \cite{n21,n22}.}. 
In particular, we demonstrate that \eqref{zeta} is valid in the vicinity of the phase transition with 
vanishing speed of sound \cite{cr1,cr2,cr3}. Notice that in this latter case  the finiteness of the 
bulk viscosity at criticality, along with the vanishing of the speed of sound at criticality, implies that 
the derivatives of the scalar fields must diverge at the critical point for \eqref{zeta} to be valid.  
Results are presented in sections 3.1 and 3.2.

It would be interesting to understand precisely why \eqref{zeta} fails for phenomenological models \cite{f1,f2,f3}. 
It is intriguing that these models also violate the bulk viscosity bound 
proposed in \cite{n21}. Thus, the derivation of \eqref{zeta} might be the first step towards proving 
the bulk viscosity bound in formal holographic examples. 

\section{Bulk viscosity of the cascading plasma}

Effective gravitational action for the cascading gauge theory plasma was derived in \cite{c1}
\begin{equation}
\begin{split}
S_5= \frac{1}{16\pi G_5} \int_{\calm_5} {{\rm vol}}_{\calm_5}\
 \biggl\lbrace &
R_5-\frac{40}{3}(\del f)^2-20(\del w)^2-\frac 12 (\del\Phi)^2-\frac{1}{4P^2}(\del K)^2 e^{-\Phi-4f-4w}\\
&-\calp
\biggr\rbrace\,,
\end{split}
\eqlabel{5actionE}
\end{equation}
where
\begin{equation}
\calp=-24 e^{-\ft {16}{3}f-2w}+4 e^{-\ft {16}{3}f-12w}+P^2 e^{\Phi-\ft {28}{3}f+4w}+\frac 12 K^2 e^{-\ft {40}{3} f }\,.
\eqlabel{calp}
\end{equation}
In this case \eqref{zeta} reads\footnote{We use $ _{EO}$ to distinguish the expressions obtained 
applying \eqref{zeta}.}
\begin{equation}
\begin{split}
\frac{\zeta}{\eta}\bigg|_{EO}=&c_s^4 T^2\biggl[\left(\frac{d\Phi^H}{dT}\right)^2+\frac{80}{3}
\left(\frac{df^H}{dT}\right)^2+40 \left(\frac{dw^H}{dT}\right)^2+\frac{1}{2P^2} e^{-\Phi^H-4 f^H-4 w^H }
\left(\frac{dK^H}{dT}\right)^2
\biggr]\,.
\end{split}
\eqlabel{eoc}
\end{equation}

As argued in \cite{c2}, it is technically challenging to study bulk viscosity of the cascading plasma for all 
temperatures. Instead, it was studied in \cite{c2} perturbatively (to the third sub-leading order) at high temperature.
From (2.3)-(2.7) of \cite{c2}
\begin{equation}
\begin{split}
&h(x)= \frac {K_{\star}}{4\tilde{a}_0^2}+\frac{K_\star}{\ta_0^2}\  
\sum_{n=1}^{\infty} \left\{\left(\frac{P^2}{K_\star}\right)^n
\left(\xi_{2n}(x)-\frac 54 \eta_{2n}(x)\right)\right\}\,,\\
\end{split}
\eqlabel{p2order1}
\end{equation}
\begin{equation}
\begin{split}
&f_2(x)=\tilde{a}_0+\ta_0\ \sum_{n=1}^\infty \left\{\left(\frac{P^2}{K_\star}\right)^n  \left(-2 \xi_{2n}(x)+\eta_{2n}(x)+\frac 45 \lambda_{2n}(x)\right)
\right\}\,,\\
\end{split}
\eqlabel{p2order2}
\end{equation}
\begin{equation}
\begin{split}
&f_3(x)=\tilde{a}_0+\ta_0\ \sum_{n=1}^\infty \left\{\left(\frac{P^2}{K_\star}\right)^n  \left(-2 \xi_{2n}(x)+\eta_{2n}(x)-\frac 15 \lambda_{2n}(x)\right)
\right\}\,,\\
\end{split}
\eqlabel{p2order3}
\end{equation}
\begin{equation}
\begin{split}
&K(x)=K_{\star}+K_{\star}\ \sum_{n=1}^\infty \left\{\left(\frac{P^2}{K_\star}\right)^n \k_{2n}(x)\right\}\,,\\
\end{split}
\eqlabel{p2order4}
\end{equation}
\begin{equation}
\begin{split}
&g(x)=1+\sum_{n=1}^\infty \left\{\left(\frac{P^2}{K_\star}\right)^n \zeta_{2n}(x)\right\}\,.\\
\end{split}
\eqlabel{p2order5}
\end{equation}
where 
\begin{equation}
\left\{f\equiv \frac 14 \ln h +\frac 25 \ln f_3+\frac {1}{10}\ln f_2\,,\qquad \w=\frac{1}{10}\ln\frac{f_3}{f_2}\,,\qquad \Phi=\ln g\right\}\,.
\eqlabel{defsc}
\end{equation}
Notice that $\ta_0$ dependence disappears for the scalars entering \eqref{eoc}, and the only temperature dependence 
enters via $K_\star=K_\star(T)$. We refer the reader to \cite{c2} for the asymptotic parametrization 
of various functions $\{\kappa_{2n},\xi_{2n},\eta_{2n},\lambda_{2n},\zeta_{2n}\}$.

The pressure $\calp$,  the energy density $\cale$, and the entropy density $s$ of the cascading plasma 
are given by \cite{aby}
\begin{equation}
\frac{\calp}{sT}=\frac 37\left(\frac{7}{12}-\ha_{2,0}\right)\,, \qquad 
\frac{\cale}{sT}=\frac 34 \left(1+\frac 47\ \ha_{2,0}\right)\,,
\eqlabel{pe}
\end{equation}
with (see (2.58) of \cite{c2})
\begin{equation}
\begin{split}
\ha_{2,0}=&\frac{7}{12} \frac{P^2}{K_\star}+\left(\frac 76 \k_2^{2,0}-\frac{35}{3}+\frac{7}{12} \z_1^{2,0}
+\frac{7}{24} \ln 2\right) \frac{P^4}{K_\star^2}
+\biggl(-\frac{35}{36} \ln 2-\frac{35}{18} \k_2^{2,0}+\frac{7}{48} \ln^2 2\\
&+\frac{175}{108}+\frac{7}{24} \z_1^{2,0}\ln 2 
+\frac{7}{12} \k_2^{2,0} \ln 2
+\frac{7}{12} \z_2^{2,0}+\frac 76 \k_3^{2,0}-\frac{35}{36} \z_1^{2,0}\biggr) \frac{P^6}{K_\star^3}+\biggl(
\frac{175}{72} \ln 2\\
&+\frac{7}{12} \z_3^{2,0}+\frac{175}{54} \k_2^{2,0}-\frac{35}{18} \k_2^{2,0} \ln 2-\frac{35}{36} \z_1^{2,0}\ln 2 
-\frac{875}{324}+\frac{175}{108} \z_1^{2,0}-\frac{35}{48} \ln^2 2\\
&+\frac{7}{96} \ln^3 2+\frac{7}{48}  \z_1^{2,0}\ln^2 2+\frac{7}{12} \k_3^{2,0}\ln 2 +\frac{7}{24}  \k_2^{2,0}\ln^2 2
+\frac{7}{24}  \z_2^{2,0}\ln 2+\frac 76 \k_4^{2,0}\\
&-\frac{35}{36} \z_2^{2,0}-\frac{35}{18} \k_3^{2,0}\biggr) 
\frac{P^8}{K_\star^4}+\calo\left(\frac{P^{10}}{K_\star^5}\right)\,.
\end{split}
\eqlabel{hadef}
\end{equation}
The precise temperature dependence of $K_\star$ was determined in \cite{kt3}
\begin{equation}
\frac{K_\star}{P^2}=\frac 12 \ln \left(\frac{64\pi^4 }{81}\ \times\ \frac{s T }{ \Lambda^4}\right)\,.
\eqlabel{ks}
\end{equation}
Using \eqref{ks} and  the expressions for the pressure and the energy density from  
\eqref{pe}  we find 
\begin{equation}
\begin{split}
c_s^2=\frac{\del\calp}{\del\cale}=\frac 13\ \frac{7-12\ha_{2,0}
-6 P^2\ \frac{d\ha_{2,0}}{d K\star}}{7+4 \ha_{2,0}+2 P^2\ \frac{d\ha_{2,0}}{d K\star}}\,.
\end{split}
\eqlabel{cs2gt}
\end{equation}
Thus, given the perturbative high temperature expansion for $\ha_{2,0}$ we can evaluate from 
\eqref{cs2gt} the 
perturbative high temperature expansion for $c_s^2$
\begin{equation}
\begin{split}
3 c_s^2=&=1-\frac 43 \frac{P^2}{K_\star}+\left(\frac{10}{3}
-\frac 23 \ln 2-\frac 83 \k_2^{2,0}-\frac 43 \z_1^{2,0}\right) \frac{P^4}{K_\star^2}+\biggl(
-8-\frac 23  \z_1^{2,0}\ln 2\\
&+\frac{10}{3} \ln 2-\frac 13 \ln^2 2-\frac 83 \k_3^{2,0}+\frac{40}{9} \z_1^{2,0}-\frac 43 \k_2^{2,0} \ln 2
+\frac{80}{9} \k_2^{2,0}-\frac 43 \z_2^{2,0}\biggr) \frac{P^6}{K_\star^3}\\
&+\biggl(-\frac 13 \ln^2 2 \z_1^{2,0}-\frac 43 \k_3^{2,0} \ln 2
-\frac 23 \k_2^{2,0}\ln^2 2 -\frac 23 \z_2^{2,0} \ln 2 +\frac{16}{9} (\k_2^{2,0})^2+\frac{16}{9} \k_2^{2,0} \z_1^{2,0}
\\
&+\frac 49 (\z_1^{2,0})^2-\frac 16 \ln^3 2-12 \ln 2+\frac{37}{9}  \z_1^{2,0}\ln 2+\frac{169}{9}+\frac{74}{9} \k_2^{2,0} \ln 2
+\frac 52 \ln^2 2\\
&-\frac{212}{9} \k_2^{2,0}-\frac{106}{9} \z_1^{2,0}+\frac{46}{9} \z_2^{2,0}+\frac{92}{9} \k_3^{2,0}
-\frac 43 \z_3^{2,0}-\frac 83 \k_4^{2,0}\biggr) \frac{P^8}{K_\star^4}+\calo\left(\frac{P^{10}}{K_\star^5}\right)\,.
\end{split}
\eqlabel{cs2eos}
\end{equation}

Consistency of the first law of thermodynamics (which was verified in \cite{c2} with an 
accuracy of $\propto 10^{-7}$) implies (see (4.5) of \cite{c2})
\begin{equation}
\begin{split}
&\frac{dK_\star}{d\ln T}=\\
&2 P^2+\left(-\frac 43+2 \z_1^{2,0}+4 \k_2^{2,0}+\ln 2\right) \frac{P^4}{K_\star}
+\biggl(4 \k_3^{2,0}-\frac 83 \k_2^{2,0}-\frac 85 (\l_{1,h}^0)^2
-\frac43 \z_1^{2,0}+\frac{79}{18}\\
&-60 (\eta_{1,h}^0)^2+2 \z_2^{2,0}+2 \k_2^{2,0} \ln 2
+\frac 12 \ln^2 2-3 \ln 2
+\ln 2\ \z_1^{2,0}-24 \xi_{2,h}^0\biggr) \frac{P^6}{K_\star^2}
\\
&+\biggl(\frac{55}{9} \ln 2+\frac{91}{18} \z_1^{2,0}+\frac{91}{9} \k_2^{2,0}-\frac 73 \z_2^{2,0}-\frac {14}{3} \k_{3}^{2,0}
+2 \z_3^{2,0}+4 \k_4^{2,0}+16 \xi_{2,h}^0-\frac{7}{2} \ln 2\ \z_1^{2,0}\\
&-7 \k_2^{2,0} \ln 2-\frac 94 \ln^2 2+\ln 2\
 \z_2^{2,0}+\frac12 \ln^2 2\ \z_1^{2,0}+2 \ln 2\ \k_3^{2,0}+\ln^2 2\ \k_2^{2,0}+\frac14 \ln^3 2
\\
&+40 (\eta_{1,h}^0)^2+\frac{16}{15} (\l_{1,h}^0)^2-36 \xi_{3,h}^0-180 \eta_{1,h}^0\ \eta_{2,h}^0+\frac{24}{5} \eta_{1,h}^0\
 (\l_{1,h}^0)^2-\frac{24}{5} \l_{1,h}^0\ \l_{2,h}^0\\
&-240 (\eta_{1,h}^0)^3+\frac{24}{25} (\l_{1,h}^0)^3+180 \xi_{2,h}^0\ \eta_{1,h}^0
-30 \ln 2\ (\eta_{1,h}^0)^2-12 \ln 2\ \xi_{2,h}^0-\frac 45 \ln 2\ (\l_{1,h}^0)^2\\
&-\frac{607}{108}-120 (\eta_{1,h}^0)^2\ \k_2^{2,0}
-60 (\eta_{1,h}^0)^2\ \z_1^{2,0}-48 \xi_{2,h}^0\ \k_2^{2,0}-24 \xi_{2,h}^0\ \z_1^{2,0}-\frac{16}{5} (\l_{1,h}^0)^2\ \k_2^{2,0}
\\
&-\frac 85 (\l_{1,h}^0)^2\ \z_1^{2,0}\biggr) \frac{P^8}{K_\star^3}+\calo\left(\frac{P^{10}}{K_\star^4}\right)\,.
\end{split}
\eqlabel{dkdt2}
\end{equation}

We now have all the necessary ingredients of evaluate \eqref{eoc} to 
order $\calo\left(\frac{P^{10}}{K_\star^5}\right)$
\begin{equation}
\begin{split}
&\frac{\zeta}{\eta}\bigg|_{EO}=\frac 89\ \frac{P^2}{K_\star}
+\biggl(-\frac{76}{27}+\frac{16}{9} \z_1^{2,0}+\frac{32}{9} \k_2^{2,0}+\frac 89 \ln 2-\frac 89 \z_{1,h}^0
+\frac 83 \eta_{1,h}^0+\frac{16}{45} \l_{1,h}^0\biggr)\frac{P^4}{K_\star^2}\\
&+\biggl(12+\frac{32}{9} \k_2^{2,0} \ln 2+\frac{16}{9} \ln 2 \z_1^{2,0}+\frac{32}{9} \k_2^{2,0} \z_1^{2,0}
-\frac83 \z_{1,h}^0 \eta_{1,h}^0-\frac{16}{45} \z_{1,h}^0 \l_{1,h}^0-\frac{32}{9} \z_{1,h}^0 \k_2^{2,0}
\\
&+\frac{16}{3} \eta_{1,h}^0 \z_1^{2,0}+\frac{32}{3} \eta_{1,h}^0 \k_2^{2,0}+\frac{32}{45} \l_{1,h}^0 \z_1^{2,0}
+\frac{64}{45} \l_{1,h}^0 \k_2^{2,0}-\frac89 \z_{1,h}^0 \ln 2+\frac83 \eta_{1,h}^0 \ln 2
\\
&+\frac{16}{45} \l_{1,h}^0 \ln 2
-\frac{16}{9} \z_{1,h}^0 \z_1^{2,0}+\frac{32}{45} \eta_{1,h}^0 \l_{1,h}^0
-8 \z_1^{2,0}-16 \k_2^{2,0}+\frac{32}{9} \z_{1,h}^0-\frac{56}{9} \eta_{1,h}^0-\frac{64}{45} \l_{1,h}^0\\
&-\frac{16}{9}
 k_{2,h}^0-\frac{64}{3} \xi_{2,h}^0+\frac{16}{45} \l_{2,h}^0-\frac89 z_{2,h}^0+\frac{16}{9} \z_2^{2,0}+\frac{32}{9} 
\k_{3}^{2,0}
+\frac 83 \eta_{2,h}^0-\frac{164}{27} \ln 2+\frac89 (\z_{1,h}^0)^2\\
&+\frac23 (\ln 2)^2+\frac{32}{9} (\k_2^{2,0})^2
+\frac89 (\z_1^{2,0})^2-\frac{112}{3} (\eta_{1,h}^0)^2-\frac{296}{225} (\l_{1,h}^0)^2\biggr)\frac{P^6}{K_\star^3}\\
&+\calv_4\ \frac{P^8}{K_\star^4}+\calo\left(\frac{P^{10}}{K_\star^5}\right)\,,
\end{split}
\eqlabel{reseoc}
\end{equation}
where the coefficient $\calv_4$ is given in Appendix.

Eq.~\eqref{reseoc} should be compared with the expression for the bulk viscosity derived in 
\cite{c2} (see (4.14) in \cite{c2})
\begin{equation}
\frac{\zeta}{\eta}=\frac 89\ \frac{P^2}{K_\star}+\frac 43\left(
\b_{2,2}\ \frac{P^4}{K_\star^2}+\b_{2,3}\ \frac{P^6}{K_\star^3}+\b_{2,4}\ \frac{P^8}{K_\star^4}
\right)+\calo\left(\frac{P^{10}}{K_\star^5}\right)\,.
\eqlabel{ze1}
\end{equation}

For convenience we collect in Table 1 numerical values for all the coefficients entering \eqref{reseoc} and \eqref{ze1} 
as computed in \cite{c2}.
\begin{table}
\centerline{
\\
\begin{tabular}
{||c||c|c|c|c||}
	\hline
\textbf{\em n}  &  $1$    &   $2$  & $3$  & $4$\\
\hline
\hline
$\k_{n}^{2,0}$ &  &0.73675974 & -0.62226255&-0.03784377\\
$\eta_{n}^{4,0}$ &-0.01717287  &0.00534036 &-0.01064222 &\\
$\l_{n}^{3,0}$ & -0.87235794 &-1.11562943 & 1.39008636&\\
$\z_{n}^{2,0}$ & -0.15342641 &0.62226267 & -0.32514260&\\
$\k_{n,h}^{0}$ &  & 0.62226259& -0.42061461&0.00816831\\
$\xi_{n,h}^{0}$ &  &-0.07981931 & 0.01661150&-0.00920379\\
$\xi_{n,h}^{0}$ &  &0.01919989 & -0.05277626&0.01385333\\
$\eta_{n,h}^{0}$ &-0.14891337  &-0.21809464 & 0.00213345&\\
$\l_{n,h}^{0}$ &  0.16806881& -0.14619173& 0.01639579&\\
$\z_{n,h}^{0}$ & -0.41123352 &0.33024116 & -0.07445122&\\
$\b_{2,n}$  &  & 0.13225837& -1.69770959 & 2.26988336\\
\hline
\end{tabular}
}
\caption{Coefficients entering \eqref{reseoc} and \eqref{ze1}.}
\label{table1}
\end{table}

\subsection{Comparison of \eqref{zeta} with bulk viscosity of the cascading plasma}

Using \eqref{reseoc} and \eqref{ze1} and the data from Table 1 we find:
\begin{equation}
\begin{split}
\frac{\zeta}{\eta}\bigg|_{EO}=&
\frac 89 \frac{P^2}{K_\star}
+.1763445167 \frac{P^4}{K_\star^2}-2.263612386 \frac{P^6}{K_\star^3}+3.026509967 \frac{P^8}{K_\star^4}
+\calo\left(\frac{P^{10}}{K_\star^5}\right)\,,
\\
\frac{\zeta}{\eta}=&\frac 89 \frac{P^2}{K_\star}+.1763444983 \frac{P^4}{K_\star^2}-2.2636127883 \frac{P^6}{K_\star^2}
+3.026511143 \frac{P^8}{K_\star^4}+\calo\left(\frac{P^{10}}{K_\star^5}\right)\,.
\end{split}
\eqlabel{cacomp}
\end{equation}

\section{Bulk viscosity of $\caln=2^*$ plasma}
Effective gravitational action for $\caln=2^*$ gauge theory plasma 
was obtained in \cite{pw}
\begin{equation}
\begin{split}
S=&\frac{1}{4\pi G_5}\,
\int_{\calm_5} d\xi^5 \sqrt{-g}\left[\ft14 R-3 (\del\a)^2-(\del\chi)^2-
\calp\right]\,,
\end{split}
\eqlabel{action5}
\end{equation}
where the potential%
\footnote{We set the five-dimensional gauged
supergravity coupling to one. This corresponds to setting the
radius $L$ of the five-dimensional sphere in the undeformed metric
to $2$.}
\begin{equation}
\calp=\frac{1}{16}\left[\frac 13 \left(\frac{\del W}{\del
\a}\right)^2+ \left(\frac{\del W}{\del \chi}\right)^2\right]-\frac
13 W^2\,
 \eqlabel{pp}
\end{equation}
is a function of $\alpha$ and $\chi$, and is determined by the
superpotential
\begin{equation}
W=- e^{-2\alpha} - \frac{1}{2} e^{4\alpha} \cosh(2\chi)\,.
\eqlabel{supp}
\end{equation}
In this case \eqref{zeta} reads:
\begin{equation}
\begin{split}
\frac{\zeta}{\eta}\bigg|_{EO}=&c_s^4 T^2\biggl[24 \left(\frac{d\a^H}{dT}\right)^2
+8 \left(\frac{d\chi^H}{dT}\right)^2
\biggr]\,.
\end{split}
\eqlabel{eon2}
\end{equation}
In what follows we present results only for  $\caln=2^*$ gauge theory plasma 
at vanishing fermionic mass\footnote{Although we did not verify this fact explicitly, 
based on the agreement for $m_f=0$, we do not expect discrepancy between \eqref{eon2} and the 
direct computation of the bulk viscosity for $m_f\ne 0$.} \cite{cr1}. 
The latter case corresponds to setting $\chi\equiv 0$.

As pointed out in \cite{cr1}, and further explored in \cite{cr2,cr3}, 
strongly coupled $\caln=2^*$ gauge theory plasma undergoes a  
second-order phase transition at 
\begin{equation}
\delta\equiv \left(\frac{m_b}{T}\right)^2=\delta_c=5.4098(6)\,.
\eqlabel{tc}
\end{equation}
At the critical point the speed of sound vanishes. 
We discuss separately the cases
$\dd<\dd_c$, and $|\dd_c-\dd|\ll 1$.

We use asymptotic parametrization of the scalar function, the speed of sound, and the bulk viscosity 
as in \cite{n22} 
\begin{equation}
\r\equiv \ln \a\,,\qquad c_s\equiv \frac{\b_1}{\sqrt{3}}\,,\qquad \frac{\zeta}{\eta}=\frac 43(\b_2-1)\,.
\eqlabel{parametrization}
\end{equation}

\subsection{Comparison of \eqref{zeta} with $\caln=2^*$ bulk viscosity away from criticality: $\dd<\dd_c$}

\begin{figure}[t]
\begin{center}
\psfrag{r0}{{$\r_0$}}
\psfrag{dd}{{$\delta$}}
\includegraphics[width=4in]{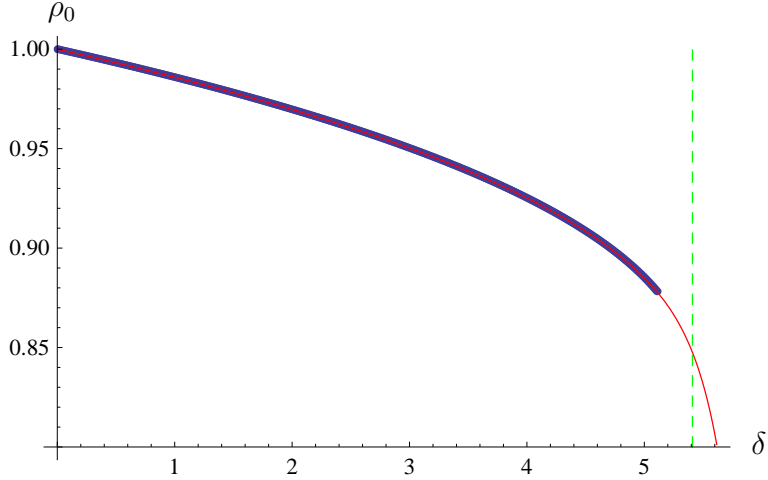}
\end{center}
  \caption{(Colour online)
Horizon value of the scalar field $\r_0=e^{\a^H}$ as the function of $\dd\equiv \left(\frac{m_b}{T}
\right)^2$. Data points are blue; the solid red line is the  best fit of the data with order 15 polynomial in 
$\dd$. The dashed vertical green line represents the critical point of the theory $\dd=\dd_c$. 
 } \label{figure1}
\end{figure}

\begin{figure}[t]
\begin{center}
\psfrag{zz}{{$\left(\frac{{\zeta}/{\eta}|_{EO}}{{\zeta}/{\eta}}-1\right)$}}
\psfrag{dd}{{$\delta$}}
\includegraphics[width=4in]{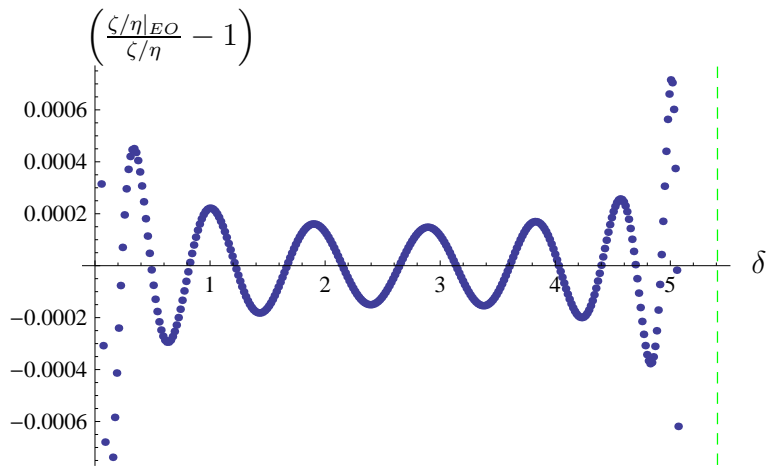}
\end{center}
  \caption{(Colour online)
Comparison of the EO prediction for  $\caln=2^*$ plasma bulk viscosity with the explicit 
computations from the quasinormal modes \cite{n21}.   
The dashed vertical green line represents the critical point of the theory $\dd=\dd_c$.
 } \label{figure2}
\end{figure}

We rewrite \eqref{eon2} assuming functional dependence $\r^H\equiv \ln\a^H=\r_0(\dd)$
\begin{equation}
\frac{\zeta}{\eta}\bigg|_{EO}=\frac{32\dd^2\b_1^4}{3\r_0^2}\left(\frac{d\r_0}{d\dd}\right)^2\,.
\eqlabel{eon21}
\end{equation}
From eq. (2.11) of \cite{n22}, 
\begin{equation}
\dd=\frac{24\pi^2}{\sqrt{2}}\ \r_{11} e^{6a_0}\,.
\eqlabel{defdd}
\end{equation}
Thus, given the data sets $\{\r_{11},\r_{10},\r_0,a_0,a_1,\b_1,\b_2\}$ 
obtained in \cite{n21,n22} we can reconstruct the functional dependence $\r_0(\dd)$. 
The result of such reconstruction is presented in Figure 1. Data points are blue ---
there are altogether 320 points. The solid red line is the best fit to the data 
with the polynomial
\begin{equation}
\r_0^{fit}=\sum_{i=0}^{15} \calr_i\ \dd^i\,.
\eqlabel{fit}
\end{equation}
Having an (approximate) analytic expression for $\r_0\approx \r_0^{fit}(\dd)$, we can compute the 
prediction for the bulk viscosity of $\caln=2^*$ plasma from \eqref{zeta}  
\begin{equation}
\frac{\zeta}{\eta}\bigg|_{EO}=\frac{32\dd^2\b_1^4}{3\r_0^2}\left(\frac{d\r_0^{fit}}{d\dd}\right)^2\,.
\eqlabel{eon22}
\end{equation}
We find that the numerical agreement between \eqref{eon22} and  the original result \eqref{parametrization}
is excellent --- Figure 2 presents the functional dependence of 
\begin{equation}
\left(\frac{{\zeta}/{\eta}|_{EO}}{{\zeta}/{\eta}}-1\right)\qquad {\rm vs}\qquad \dd\,.
\eqlabel{resn2}
\end{equation}
Again, the vertical green dashed line represents the critical point of $\caln=2^*$ plasma.
The agreement is relatively worse for small values of $\dd$ and in the vicinity of the 
critical point. In the former case, this is caused by numerical errors for the evaluation of the 
bulk viscosity ( the larger errors are  induced due  to small amplitude profiles of the gravitational scalar 
field $\a$ ),
and in the latter case the  relatively worse agreement is caused by the choice of the 'fit' 
\eqref{fit} (in the vicinity of the critical point $\frac{d\r_0}{d\dd}$ is expected to diverge as $\b_1^{-2}$).  
As was already pointed out in \cite{eo}, the agreement between  \eqref{eon21} and \eqref{parametrization}
can be established analytically to order $\calo(\dd)$. In the next subsection, with a suitable 
parametrization of the horizon value of the scalar field at criticality, we drastically improve 
the agreement  between  \eqref{eon21} and \eqref{parametrization} for $\dd$ close to $\dd_c$.

\subsection{Comparison of \eqref{zeta} with $\caln=2^*$ bulk viscosity at criticality  $|\dd_c-\dd|\ll 1$}

\begin{figure}[t]
\begin{center}
\psfrag{r0}{{$\r_0$}}
\psfrag{dd}{{$\delta$}}
\psfrag{beta}{{$\b$}}
\includegraphics[width=3in]{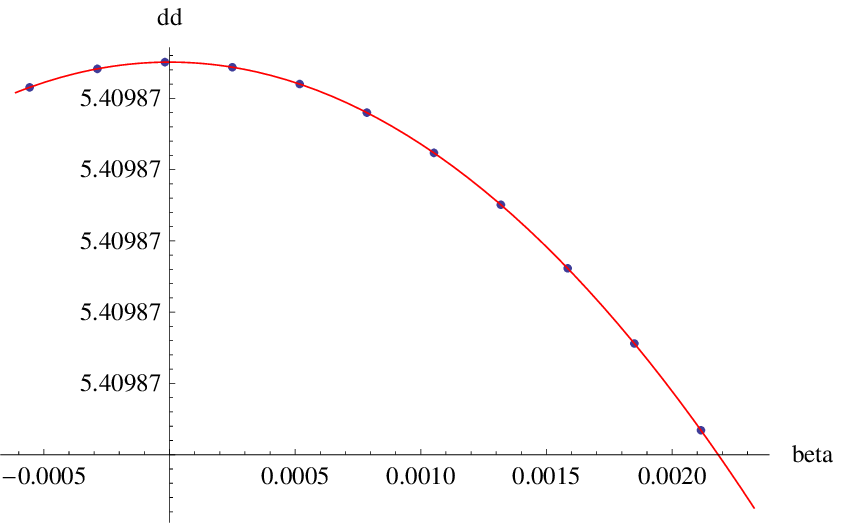}
\includegraphics[width=3in]{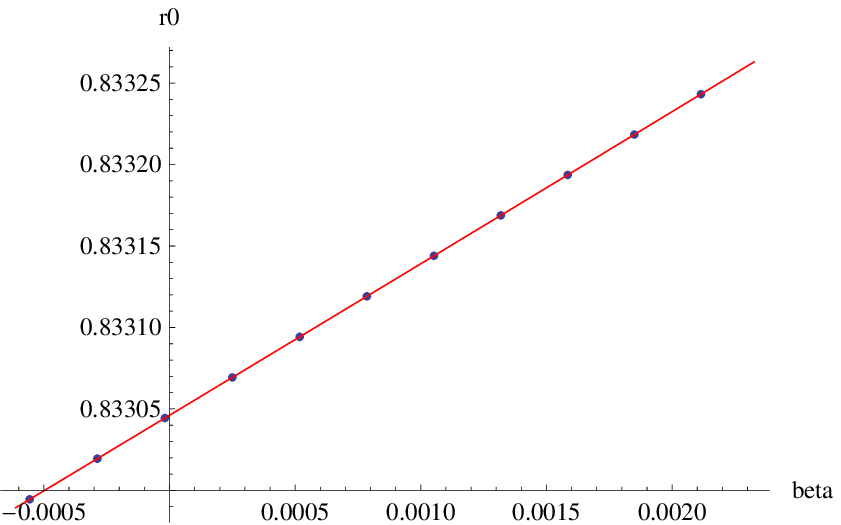}
\end{center}
  \caption{(Colour online)
Horizon value of the scalar field $\r_0=e^{\a^H}$ and  $\dd\equiv \left(\frac{m_b}{T}
\right)^2$ as a function of $\b\equiv c_s^2$ at criticality. Data points are blue; the solid red lines 
are the  best fit of the data with  polynomials in 
$\b$ --- see \eqref{fitf}. 
 } \label{figure3}
\end{figure}

\begin{figure}[t]
\begin{center}
\psfrag{zz}{{$\left(\frac{{\zeta}/{\eta}|_{EO}}{{\zeta}/{\eta}}-1\right)$}}
\psfrag{beta}{{$\beta$}}
\includegraphics[width=4in]{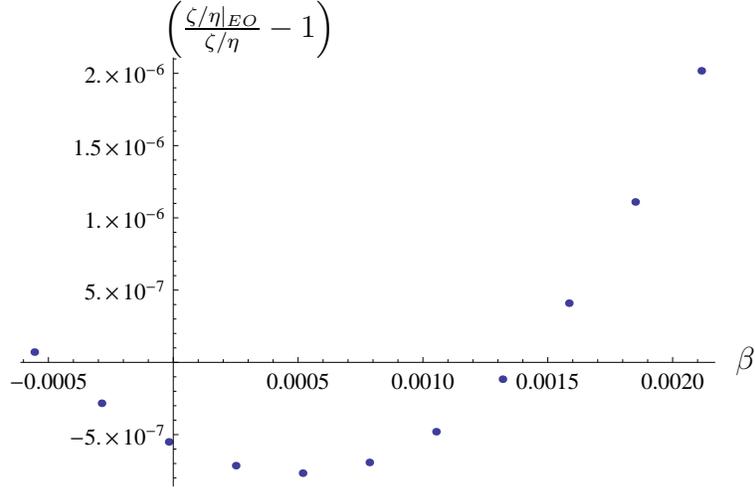}
\end{center}
  \caption{(Colour online)
Comparison of the EO prediction for  $\caln=2^*$ plasma bulk viscosity with the explicit 
computations from the quasinormal modes \cite{n21,n22} in the vicinity of the critical point,
\ie, for $\b=0$.
 } \label{figure4}
\end{figure}

An important  prediction of \cite{n21,n22} is that the bulk viscosity of  $\caln=2^*$ plasma remains 
finite at criticality, where $c_s^2=0$. The EO expression \eqref{eon21}, if correct, implies that 
the derivative of the scalar field $\r_0$ with respect to $\dd$ must diverge as $c_s^{-2}$ at criticality. 
Thus $\dd$ parametrization of $\r_0$ is not very useful. Instead, in the vicinity of the 
critical point we rewrite \eqref{eon21} as a function of 
\begin{equation}
\beta\equiv c_s^2\equiv \frac{\b_1^2}{3}\,.
\eqlabel{beta}
\end{equation}
We find
\begin{equation}
\frac{\zeta}{\eta}\bigg|_{EO}=\frac{96\dd^2}{\r_0^2}\left(\frac{d\r_0}{d\b}\right)^2\left(\beta\ \frac{d\b}{d\dd}\right)^2\,,
\eqlabel{eon23}
\end{equation}
where now we understand $\r_0=\r_0(\b)$ and $\dd=\dd(\b)$. These two functions can easily be reconstructed 
from the data sets  $\{\r_{11},\r_{10},\r_0,a_0,a_1,\b_1,\b_2\}$ 
obtained in \cite{n21,n22} --- see Figure 3. Blue dots represent the data points, the solid red lines are the 
fits to the data:
\begin{equation}
\begin{split}
\dd^{fit}=&5.40987 - 1.14476\ \b^2 - 4.64346\ \b^3 -14.55312\ \b^4\,,\\
\r_0^{fit}=&0.83305+0.09281\ \b+0.17683\ \b^2+0.36473\ \b^3\,.
\end{split}
\eqlabel{fitf}
\end{equation}
We can now compute the 
prediction for the bulk viscosity of $\caln=2^*$ plasma from \eqref{zeta}  
\begin{equation}
\frac{\zeta}{\eta}\bigg|_{EO}=\frac{96\dd^2}{\r_0^2}\left(\frac{d\r_0^{fit}}{d\b}\right)^2\left(\beta\ \frac{d\b}
{d\dd^{fit}}\right)^2\,.
\eqlabel{eon24}
\end{equation}
Figure 4 presents the functional dependence of 
\begin{equation}
\left(\frac{{\zeta}/{\eta}|_{EO}}{{\zeta}/{\eta}}-1\right)\qquad {\rm vs}\qquad \b\,,
\eqlabel{resn23}
\end{equation}
in the vicinity of the critical point of  $\caln=2^*$ plasma.

\section*{Acknowledgments}
I would like to thank Yaron Oz for valuable correspondence.
Research at Perimeter Institute is
supported by the Government of Canada through Industry Canada and by
the Province of Ontario through the Ministry of Research \&
Innovation. AB gratefully acknowledges further support by an NSERC
Discovery grant and support through the Early Researcher Award
program by the Province of Ontario. 

\appendix
\section{Coefficient $\calv_4$}

\begin{equation}
\begin{split}
&\calv_4=\frac{3080}{81} \z_1^{2,0}+\frac{6160}{81} \k_2^{2,0}-\frac{404}{27} \z_{1,h}^0+\frac{244}{9} \eta_{1,h}^0
+\frac{808}{135} \l_{1,h}^0+\frac{64}{9} \k_{2,h}^0+\frac{608}{9} \xi_{2,h}^0-\frac{64}{45} \l_{2,h}^0
\\
&+\frac{32}{9} \z_{2,h}^0-\frac{80}{9} \z_2^{2,0}-\frac{160}{9} \k_3^{2,0}-\frac{16}{9} \eta_{2,h}^0
-\frac{32}{9} \k_{3,h}^0-32 \xi_{3,h}^0+\frac 83 \eta_{3,h}^0+\frac{16}{45} \l_{3,h}^0-\frac89
 \z_{3,h}^0\\
&+\frac{32}{9} \k_4^{2,0}+\frac{16}{9} \z_3^{2,0}-\frac{32}{45} \z_{1,h}^0 \eta_{1,h}^0 \l_{1,h}^0
+\frac{64}{45} \eta_{1,h}^0 \l_{1,h}^0 \z_1^{2,0}+\frac{128}{45} \eta_{1,h}^0 \l_{1,h}^0 \k_2^{2,0}-\frac{16}{3}
 \z_{1,h}^0 \eta_{1,h}^0 \z_1^{2,0}\\
&-\frac{32}{3} \z_{1,h}^0 \eta_{1,h}^0 \k_2^{2,0}-\frac{32}{45} \z_{1,h}^0 \l_{1,h}^0 \z_1^{2,0}
-\frac{64}{45} \z_{1,h}^0 \l_{1,h}^0 \k_2^{2,0}-\frac{32}{9} \z_{1,h}^0 \k_2^{2,0} \z_1^{2,0}+\frac{32}{3} 
\eta_{1,h}^0 \k_2^{2,0} \z_1^{2,0}\\
&+\frac{64}{45} \l_{1,h}^0 \k_2^{2,0} \z_1^{2,0}-\frac{272}{9} \k_2^{2,0} \z_1^{2,0}
+\frac{32}{3} \z_{1,h}^0 \eta_{1,h}^0+\frac{64}{45} \z_{1,h}^0 \l_{1,h}^0+\frac{512}{27} \z_{1,h}^0 \k_2^{2,0}
-\frac{176}{9} \eta_{1,h}^0 \z_1^{2,0}\\
&-\frac{352}{9} \eta_{1,h}^0 \k_2^{2,0}-\frac{512}{135} \l_{1,h}^0 \z_1^{2,0}
-\frac{1024}{135} \l_{1,h}^0 \k_2^{2,0}+\frac{256}{27} \z_{1,h}^0 \z_1^{2,0}-\frac{128}{45} \eta_{1,h}^0 \l_{1,h}^0
\\
&-\frac{704}{45} (\eta_{1,h}^0)^2 \l_{1,h}^0
+\frac{8}{75} \eta_{1,h}^0 (\l_{1,h}^0)^2+\frac{16}{9} \z_{1,h}^0 \z_{2,h}^0
-\frac{448}{3} (\eta_{1,h}^0)^2 \k_2^{2,0}-\frac{224}{3} (\eta_{1,h}^0)^2 \z_1^{2,0}\\
&-\frac{256}{3} \xi_{2,h}^0 \k_2^{2,0}
-\frac{128}{3} \xi_{2,h}^0 \z_1^{2,0}-\frac{1184}{225} (\l_{1,h}^0)^2 \k_2^{2,0}-\frac{592}{225} (\l_{1,h}^0)^2 \z_1^{2,0}
+\frac{112}{3} \z_{1,h}^0 (\eta_{1,h}^0)^2\\
&+\frac{296}{225} \z_{1,h}^0 (\l_{1,h}^0)^2-\frac 83 \z_{1,h}^0 \eta_{2,h}^0
-\frac{16}{45} \z_{1,h}^0 \l_{2,h}^0-\frac 83 \eta_{1,h}^0 \z_{2,h}^0+\frac 83 \eta_{1,h}^0 (\z_{1,h}^0)^2-\frac{16}{45}
 \l_{1,h}^0 \z_{2,h}^0\\
&+\frac{16}{45} \l_{1,h}^0 (\z_{1,h}^0)^2-\frac{32}{9} \k_{2,h}^0 \z_1^{2,0}-\frac{64}{9}
 \k_{2,h}^0 \k_2^{2,0}+\frac{32}{9} \z_1^{2,0} \k_3^{2,0}+\frac{16}{9} \z_1^{2,0} \z_2^{2,0}+\frac{64}{9} \k_2^{2,0} 
\k_3^{2,0}\\
&+\frac{32}{9} \k_2^{2,0} \z_2^{2,0}-\frac{16}{9} \z_{2,h}^0 \z_1^{2,0}-\frac{32}{9} \z_{2,h}^0 \k_2^{2,0}
+\frac{16}{9} (\z_{1,h}^0)^2 \z_1^{2,0}+\frac{32}{9} (\z_{1,h}^0)^2 \k_2^{2,0}
+\frac{16}{3} \eta_{2,h}^0 \z_1^{2,0}\\
&+\frac{32}{3} \eta_{2,h}^0 \k_2^{2,0}+\frac{32}{45} \l_{2,h}^0 \z_1^{2,0}
+\frac{64}{45} \l_{2,h}^0 \k_2^{2,0}-\frac{32}{9} \z_{1,h}^0 \k_3^{2,0}-\frac{16}{9} \z_{1,h}^0 \z_2^{2,0}
+\frac{64}{3} \z_{1,h}^0 \xi_{2,h}^0\\
&+\frac{16}{9} \z_{1,h}^0 \k_{2,h}^0-\frac 89 \z_{1,h}^0 (\z_1^{2,0})^2-\frac{32}{9}
 \z_{1,h}^0 (\k_2^{2,0})^2+\frac{32}{3} \eta_{1,h}^0 \k_3^{2,0}+\frac{16}{3} \eta_{1,h}^0 \z_2^{2,0}-\frac{16}{3} 
\eta_{1,h}^0 \k_{2,h}^0\\
&+\frac 83 \eta_{1,h}^0 (\z_1^{2,0})^2+\frac {32}{3} \eta_{1,h}^0 (\k_2^{2,0})^2+\frac{64}{45}
 \l_{1,h}^0 \k_3^{2,0}+\frac{32}{45} \l_{1,h}^0 \z_2^{2,0}-\frac{32}{45} \l_{1,h}^0 \k_{2,h}^0
\\
&+\frac{16}{45} \l_{1,h}^0 (\z_1^{2,0})^2+\frac{64}{45} \l_{1,h}^0 (\k_2^{2,0})^2+\frac{224}{3} \eta_{1,h}^0 \xi_{2,h}^0
-128 \eta_{1,h}^0 \eta_{2,h}^0+\frac{32}{45} \eta_{1,h}^0 \l_{2,h}^0-\frac{352}{45} \l_{1,h}^0 \xi_{2,h}^0\\
&+\frac{32}{45}
 \l_{1,h}^0 \eta_{2,h}^0-\frac{304}{75} \l_{1,h}^0 \l_{2,h}^0\frac{3296}{8}-\frac{28}{9} (\z_{1,h}^0)^2
-\frac{272}{9} (\k_2^{2,0})^2-\frac{68}{9} (\z_1^{2,0})^2+\frac{1364}{9} (\eta_{1,h}^0)^2\\
&+\frac{1064}{225}
 (\l_{1,h}^0)^2-\frac{2672}{9} (\eta_{1,h}^0)^3+\frac{352}{1125} (\l_{1,h}^0)^3-\frac 89 (\z_{1,h}^0)^3
+\biggl(-\frac{436}{27} \z_1^{2,0}-\frac{872}{27} \k_2^{2,0}\\
&+\frac{184}{27} \z_{1,h}^0-16 \eta_{1,h}^0
-\frac{368}{135} \l_{1,h}^0-\frac{16}{9} \k_{2,h}^0-\frac{64}{3} \xi_{2,h}^0+\frac{16}{45} \l_{2,h}^0
-\frac 89 \z_{2,h}^0+\frac{16}{9} \z_2^{2,0}+\frac{32}{9} \k_3^{2,0}\\
&+\frac 83 \eta_{2,h}^0+\frac{32}{9}
 \k_2^{2,0} \z_1^{2,0}-\frac83 \z_{1,h}^0 \eta_{1,h}^0-\frac{16}{45} \z_{1,h}^0 \l_{1,h}^0-\frac{32}{9} 
\z_{1,h}^0 \k_2^{2,0}+\frac{16}{3} \eta_{1,h}^0 \z_1^{2,0}+\frac{32}{3} \eta_{1,h}^0 \k_2^{2,0}\\
&+\frac{32}{45}
 \l_{1,h}^0 \z_1^{2,0}+\frac{64}{45} \l_{1,h}^0 \k_2^{2,0}-\frac{16}{9} \z_{1,h}^0 \z_1^{2,0}+\frac{32}{45}
 \eta_{1,h}^0 \l_{1,h}^0+\frac{2378}{81}+\frac 89 (\z_{1,h}^0)^2+\frac{32}{9} (\k_2^{2,0})^2\\
&+\frac89
 (\z_1^{2,0})^2-\frac{112}{3} (\eta_{1,h}^0)^2-\frac{296}{225} (\l_{1,h}^0)^2\biggr) \ln 2
+\biggl(\frac43 \z_1^{2,0}-\frac{179}{27}+\frac83 \k_2^{2,0}+2 \eta_{1,h}^0+\frac{4}{15} \l_{1,h}^0\\
&-\frac 23
 \z_{1,h}^0\biggr) \ln^2 2+\frac 49 \ln^3 2
\end{split}
\eqlabel{v4}
\end{equation}

\end{document}